\documentclass[12pt]{article}
\usepackage{amsmath}
\usepackage{graphicx}
\usepackage{natbib}
\usepackage{url} 
\usepackage{algorithm}
\usepackage{algpseudocode}
\usepackage{amssymb}
\usepackage{hyperref} 
\usepackage{amsthm}
\usepackage{color}
\newtheorem{theorem}{Theorem}

\newtheorem{proposition}{Proposition}
\newcommand\NoThen{\renewcommand\algorithmicthen{}}

\DeclareMathOperator*{\argmin}{argmin}
\DeclareMathOperator*{\argmax}{argmax}
\newcommand{\blind}{0}

\usepackage{ulem}

\begin{document}

\def\spacingset#1{\renewcommand{\baselinestretch}%
{#1}\small\normalsize} \spacingset{1}


\if0\blind
{
  \title{\bf Boosting Prediction with Data Missing Not at Random}
  \author{Yuan Bian\\
    Department of Statistical and Actuarial Sciences \\University of Western Ontario\\
    and \\
    Grace Y. Yi \\
    Department of Statistical and Actuarial Sciences \\Department of Computer Science \\University of Western Ontario\\
    and \\
    Wenqing He\thanks{
    corresponding author: \url{whe@stats.uwo.ca}} \\
    Department of Statistical and Actuarial Sciences \\University of Western Ontario}  
    \date{}
  \maketitle
} \fi

\if1\blind
{
  \bigskip
  \bigskip
  \bigskip
  \begin{center}
    {\LARGE\bf Boosting Prediction with Data Missing Not at Random}
\end{center}
  \medskip
} \fi

\bigskip
\begin{abstract}
Boosting has emerged as a useful machine learning technique over the past three decades, attracting increased attention. Most advancements in this area, however, have primarily focused on numerical implementation procedures, often lacking rigorous theoretical justifications. Moreover, these approaches are generally designed for datasets with fully observed data, and their validity can be compromised by the presence of missing observations. In this paper, we employ semiparametric estimation approaches to develop boosting prediction methods for data with missing responses. We explore two strategies for adjusting the loss functions to account for missingness effects. The proposed methods are implemented using a functional gradient descent algorithm, and their theoretical properties, including algorithm convergence and estimator consistency, are rigorously established. Numerical studies demonstrate that the proposed methods perform well in finite sample settings.
\end{abstract}

\noindent%
\vfill

\newpage
\spacingset{1.75} 
\section{Introduction}
Boosting, a useful machine learning method, transforms weak learners into strong learners through iterative processes \citep{SchapireFreund2012}. The \textit{AdaBoost} algorithm \citep{FreundSchapire1997} is widely recognized as the first practically feasible boosting algorithm, regarded as the best off-the-shelf classifier \citep{Breiman1998}. \cite{Breiman1998, Breiman1999} demonstrated that the AdaBoost algorithm can be interpreted as a steepest descent algorithm in a function space spanned by weak learners. \cite{FriedmanHastie2000} and \cite{Friedman2001} developed a general statistical framework, providing a direct interpretation of boosting as a function estimation method. They also extended boosting from binary classification to regression and multiclass classification. \cite{BuhlmannYu2003} introduced a computationally simple boosting algorithm for regression and classification problems using the $L_2$ loss function.

These boosting methods typically require access to complete data without missing values, which is, however, often not true in practice. Recognizing this limitation, \cite{HothornBuhlmann2006}, \cite{BarnwalCho2022}, \cite{ChenYi2024}, and \cite{BianYi2025b}, among others, extended boosting algorithms to accommodate censored response variables. In this paper, we focus on extending boosting algorithms to handle incomplete data scenarios where the response variable is subject to missingness.

When data are {\it missing completely at random} (MCAR), conventional boosting algorithms can be directly applied to the complete observations, yielding valid results as they still constitute a random subsample. However, directly applying boosting procedures to data {\it missing not at random} (MNAR) poses a significant challenge, as it can lead to biased results \citep{BianYi2025a}. This issue is further complicated by the inherent non-identifiability problems associated with MNAR data, as discussed by \cite{WangShao2014}, \cite{MiaoTchetgen2016}, \cite{SunLiu2018}, and \cite{MorikawaKim2021}, among others. 

In this paper, we incorporate two strategies to adjust the loss function: the Buckley-James-type and the inverse propensity weight adjustments, with the missingness effects accounted for. These adjustments allow the flexible application of boosting methods to incomplete data with MNAR through a functional gradient descent algorithm. However, the usual optimization procedures are complicated by the involvement of unknown functions. To address this difficulty, we describe semiparametric optimal estimation approaches that provide consistent estimators for these unknown functions. We rigorously establish the theoretical properties for the resultant estimators.  In addition, we adapt conditions considered by  \cite{MorikawaKim2021} to ensure model identifiability.

The remainder of the paper is organized as follows. Section \ref{sec: cbecd} introduces conventional boosting prediction with full data. Section \ref{sec: ubemnar} addresses incomplete data with MNAR responses and presents boosting prediction approaches that accommodate the missingness effects. Section \ref{sec: id} describes semiparametric approaches for estimating the unknown functions involved. Section \ref{sec: tr} establishes the theoretical results, followed by simulation studies in Section \ref{sec: ss}. In Section \ref{sec: da}, we analyze a real dataset to illustrate the use of the proposed methods. We conclude the article with discussions in Section \ref{sec: d} and defer technical details to the Supplementary Materials.

\section{Conventional Boosting with Full Data}
\label{sec: cbecd}
\subsection{Objective and Data}
\label{subsec: od}
Let $Y$ denote the continuous response variable, and let $X$ denote the $p$-dimensional random vector of covariates, where $Y\in\mathcal{Y}$, $X\in\mathcal{X}$, and $\mathcal{Y}$ and $\mathcal{X}$ are the sample spaces for $Y$ and $X$, respectively. The goal is to find a function of $X$, say $f(X)$, that can effectively predict $Y$. Let $\mathcal{F}=\{f: \mathcal{X}\to\mathcal{Y} \ | \ f \text{ is continuous and bounded in } \ell_\infty\}$ denote the set of real valued functions satisfying condition (B9) in the Supplementary Materials. For $f\in\mathcal{F}$, let $L:\mathcal{Y}\times \mathcal{Y} \xrightarrow{\ } \mathbb{R}$ denote the \textit{loss function} that describes the discrepancy of using $f(X)$ to predict $Y$, which is assumed to be differentiable (almost everywhere) and convex with respect to the second argument, as specified in condition (B3) of the Supplementary Materials.

Given the loss function $L(\cdot, \cdot)$, for $f\in\mathcal{F}$, define the \textit{risk function} as
\begin{equation}
\label{eq: rf}
R(f)=E\{L\left(Y, f(X)\right)\},    
\end{equation}
where the expectation is taken with respect to the joint distribution of $X$ and $Y$. To find a function $f\in\mathcal{F}$ that predicts $Y$ well, we minimize the risk function:
\begin{equation}
\label{eq: rfmin}
f^*=\argmin_{f\in\mathcal{F}}R(f).    
\end{equation}

In practice, the joint distribution of $X$ and $Y$ is unknown, and we typically have access to only a random sample of $n$ independent observations of them, denoted by $\mathcal{O}_{\text{full}}\triangleq\{\{X_i, Y_i\}: \ i=1, \ldots, n\}$. Consequently, replacing the expectation in \eqref{eq: rf} with its empirical counterpart, we estimate $f^*$ by minimizing the \textit{empirical risk}: 
\begin{equation}\label{eq: erf}
\hat{f}_{\text{full}}=\argmin_{f\in\mathcal{F}}\left\{n^{-1}\sum^n_{i=1}L(Y_i, f(X_i))\right\}.
\end{equation}

\subsection{Conventional Boosting Prediction Procedure}
\label{subsec: cbep}
Finding $\hat{f}_{\text{full}}$ in \eqref{eq: erf} can  be achieved by employing the boosting method \citep{FreundSchapire1997}. This method, also known as the \textit{functional gradient descent algorithm} \citep[e.g.,][]{BuhlmannYu2003, BuhlmannHothorn2007, SchapireFreund2012}, minimizes the empirical risk through steepest gradient descent in a function space to iteratively improve the estimate of $\hat{f}_{\text{full}}$ \citep{Friedman2001}, where the function space is spanned by a class of constrained continuous functions, denoted by $\mathcal{C}$. At iteration $m$, given the current estimate $f^{(m)}(\cdot)$, the next estimate $f^{(m+1)}$ of $\hat{f}_{\text{full}}$ is updated by adding an increment term, $\hat{\alpha}^{(m+1)}\hat{h}^{(m+1)}(\cdot)$, where $\hat{h}^{(m+1)}(\cdot)\in\mathcal{C}$, and $\hat{\alpha}^{(m+1)}$ is a learning rate.

Specifically, let 
\begin{equation*}
\hat{h}^{(m+1)}=\argmin_{h^{(m+1)}\in\mathcal{C}}\left[n^{-1}\sum^n_{i=1}\left\{\partial L\left(Y_i, f^{(m)}(X_i)\right)h^{(m+1)}(X_i)\right\}\right]
\end{equation*}
and
\begin{equation*}
\hat{\alpha}^{(m+1)}=\argmin_{\alpha^{(m+1)}\in\mathbb{R}}\left\{n^{-1}\sum^n_{i=1}L\left(Y_i, f^{(m)}(X_i)+\alpha^{(m+1)}\hat{h}^{(m+1)}(X_i)\right)\right\},
\end{equation*}
where $\partial L\left(Y_i, f^{(m)}(X_i)\right)\triangleq\left.\frac{\partial L\left(u, v\right)}{\partial v}\right|_{u=Y_i, v=f^{(m)}(X_i)}$ for $i=1, \ldots, n$. Then, at iteration $(m + 1)$, the updated function $f^{(m+1)}(\cdot)$ is given by
\begin{equation*}
f^{(m+1)}(\cdot)=f^{(m)}(\cdot)+\hat{\alpha}^{(m+1)}\hat{h}^{(m+1)}(\cdot),
\end{equation*}
or equivalently, 
$$f^{(m+1)}(\cdot)=f^{(0)}(\cdot)+\sum^{m+1}_{j=1}\hat{\alpha}^{(j)}\hat{h}^{(j)}(\cdot).$$

The iteration procedure terminates when a specified stopping criterion is met, say after $\tilde{m}$ iterations. The final estimator of $f^*$ is then given by $\hat{f}_{\text{full}}(\cdot)\triangleq f^{(\tilde{m})}(\cdot)$. A common stopping criterion evaluates the difference in the values of $L(\cdot, \cdot)$ between successive estimates, $f^{(m)}(\cdot)$ and $f^{(m-1)}(\cdot)$. The iteration stops when this difference, measured in a certain norm (e.g., $L_1$ or $L_2$), falls below a prespecified threshold value (e.g., $10^{-6}$).

\section{Boosting Prediction  with MNAR Data}\label{sec: ubemnar}
The development in Section \ref{sec: cbecd} builds upon the assumption that a random sample of full data, $\mathcal{O}_{\text{full}}$, is available. In applications, however, this assumption is often not true. Here, we focus on the scenario where the response is subject to missingness while the covariate vector is always observed. 

For $i=1, \ldots, n$, let $R_i$ denote the response missingness indicator, which equals 1 if $Y_i$ is observed and 0 otherwise. Throughout the paper, we use lowercase letters $y_i$, $x_i$, and $r_i$ to represent realizations of $Y_i$, $X_i$, and $R_i$, respectively. The observed data form a sample, denoted as $\mathcal{O}_{\text{missing}}$, which includes $\{y_i, x_i, r_i=1\}$ or $\{x_i, r_i=0\}$ for $i=1, \ldots, n$. For simplicity, we occasionally drop the subject index $i$ from the notation.

\subsection{Missing not at Random and Identifiablity}\label{subsec: MNARI}
Let $f(y|X=x)$ denote the conditional probability density of $Y$ given $X$, and let $\pi(y,x)\triangleq \Pr(R=1|Y=y,X=x)$ denote the \textit{propensity} of observing the response, i.e., the conditional probability of observing $Y$, given $Y$ and $X$. If $\pi(y,x)$ does not depend on $Y$, the resulting missing data mechanism is called \textit{missing at random} (MAR). When the missing mechanism is \textit{missing not at random} (MNAR) or \textit{nonignorable}, the propensity $\pi(y,x)$ depends on $Y$, as well as $X$, irrespective of whether $Y$ is missing or observed. Under MNAR, non-identifiability is often a concern \citep{RobinsRitov1997}. When both $f(y|X=x)$ and $\pi(y,x)$ are left fully unspecified, the joint distribution of $Y$ and $R$, given $X$, becomes non-identifiable \citep{RobinsRitov1997}. 

To address non-identifiability issues, certain assumptions can be imposed. For example, \cite{WangShao2014} assumed the existence of a \textit{nonresponse instrumental variable} (aka a \textit{shadow variable}) \citep{MiaoTchetgen2016}, which is a component of the covariate vector $X$ that is associated with the response $Y$ but is conditional independent of the missingness indicator $R$, given $Y$ and other components of $X$. Alternatively, \cite{SunLiu2018} assumed the existence of another version of instrumental variables: a subset of the covariate vector $X$ that is independent of the response $Y$ but conditionally dependent on $R$, given $Y$ and other components of $X$. While utilizing instrumental variables can mitigate non-identifiability issues, identifying an appropriate instrumental variable can be difficult in applications. Most importantly, those conditions are not testable solely based on the observed data \citep{MorikawaKim2021}.

Instead of relying on the existence of instrumental variables, \cite{MorikawaKim2021} proposed a set of identification conditions, listed as (A1) - (A4) in the Supplementary Materials. Those conditions are testable using observed data. Even if the model does not satisfy those conditions, a \textit{doubly-normalized exponential transformation} \citep{MorikawaKim2021} can be applied to artificially make the model identifiable, albeit at the cost of sacrificing the estimator consistency. In our development here, we adopt the identification conditions of \cite{MorikawaKim2021} to emphasize our key ideas.

Imposing parametric assumptions on both $f(y|X=x)$ and $\pi(y,x)$ can help address the non-identifiability issue, but the results are often sensitive to model misspecification \citep{Kenward1998}. As a remedy, an intermediate approach may be considered, where one function is modeled parametrically while leaving the other unspecified; this is the strategy we adopt in the following development.

\subsection{Adjusting Loss Functions with MNAR Data}\label{subsec: alfmda}
The boosting prediction procedure described in Section \ref{subsec: cbep} cannot be applied directly to $\mathcal{O}_{\text{missing}}$, as the response $Y_i$ is not observed for every subject in the study. To address this, we construct a new loss function, denoted $L^*(y_i, f(x_i), r_i)$, using the observed data in the sample $\mathcal{O}_{\text{missing}}$. Following the idea of \cite{ChenYi2024},  we construct an adjusted loss function $L^*(\cdot, \cdot, \cdot)$ that maintains the same risk function as the original loss function $L(\cdot, \cdot)$, i.e., $E\{L^*(Y_i, f(X_i), R_i)\}=E\{L(Y_i, f(X_i))\}$. Therefore, minimizing the risk function of the adjusted loss function $E\{L^*(Y_i,$ $f(X_i), R_i)\}$ is equivalent to minimizing the risk function $R(f)$ defined in \eqref{eq: rf}, as if all $Y_i$ were observed.

Following \cite{HothornBuhlmann2006}, one way to adjust the loss function is through the \textit{inverse propensity weight} (IPW) scheme:
\begin{equation}
\label{eq: ipw}
L_{\text{IPW}}(y_i, f(x_i), r_i)=\frac{r_iL(y_i, f(x_i))}{\pi(y_i,x_i)},
\end{equation}
where, to ensure \eqref{eq: ipw} to be well-defined, $\pi(y,x)$ is assumed to be bounded away from 0, as stated in condition (B4) in the Supplementary Materials. The adjusted loss function \eqref{eq: ipw} uses only complete measurements and discards partial information from subjects whose responses are missing.

To maximize the use of all available measurements, we follow the idea of \cite{ChenYi2024} to construct a Buckley-James (BJ)-type adjusted loss function:
\begin{equation}
\label{eq: BJ}
L_{\text{BJ}}(y_i, f(x_i), r_i)=r_iL(y_i, f(x_i))+(1-r_i)\Psi_0(x_i),    
\end{equation}
where $\Psi_0(x)\triangleq E\{L(Y, f(X))|X=x, R=0\}$, determined by $\int L(y, f(x))f(y|X=x, R=0)dy$.

Determining $\Psi_0(x)$ requires $f(y|X=x, R=0)$, which is not available since the outcome is not observed for this subpopulation. To get around this, as in \cite{KimYu2011}, we use Bayes' rule to express
\begin{equation}
\label{eq: rKY}
f(y|X=x, R=0)=\frac{f(y|X=x, R=1)O(y,x)}{E\{O(Y,X)|X=x, R=1\}},
\text{ with } 
O(y,x)\triangleq\frac{1-\pi(y,x)}{\pi(y,x)};
\end{equation}
$f(y|X=x,R=1)$ is the conditional probability density of $Y$, given $X=x$ and $R=1$; and $E\{O(Y,X)|X=x, R=1\}$ is determined by $\int O(y,x)f(y|X=x,R=1)dy$.

While incorporating $\Psi_0(x_i)$ in \eqref{eq: BJ} facilitates contributions from subjects with missing responses, it is more restrictive than \eqref{eq: ipw}, as \eqref{eq: BJ} requires two working models $f(y|X=x, R=1)$ and $\pi(y,x)$. The following proposition justifies that the proposed adjusted loss functions accommodate the missingness effects while recovering the expectation of the original loss function. The proof is placed in the Supplementary Materials. 

\begin{proposition}\label{prop1}
The proposed adjusted loss functions \eqref{eq: ipw} and \eqref{eq: BJ} have
the same expectation as $L(Y_i,f(X_i))$. That is,
\begin{itemize}
    \item[(a)] $E\{L_{\text{IPW}}(Y_i, f(X_i), R_i)\} =  E\{L(Y_i, f(X_i) )\};$
    \item[(b)] $E\{L_{\text{BJ}}(Y_i, f(X_i), R_i)\} =  E\{L(Y_i, f(X_i))\},$
\end{itemize}
where the expectations are evaluated with respect to the joint distribution of
the associated random variables.
\end{proposition}

Proposition \ref{prop1} shows that our proposed adjusted loss functions, \eqref{eq: ipw} and \eqref{eq: BJ}, have the same expectation as the original loss function constructed using the full data $\mathcal{O}_{\text{full}}$, as if they were available. Therefore, the risk function for an adjusted loss function, \eqref{eq: ipw} or \eqref{eq: BJ}, derived from the observed incomplete data $\mathcal{O}_{\text{missing}}$, is identical to that derived from $\mathcal{O}_{\text{full}}$. Consequently, the optimization problem \eqref{eq: erf} based on $\mathcal{O}_{\text{full}}$ is now converted to the following problem, which is computed from $\mathcal{O}_{\text{missing}}$: 
\begin{equation}
\label{eq: erfal}
\hat{f}^{\text{AL}}=\argmin_{f\in\mathcal{F}}\left\{n^{-1}\sum^n_{i=1}L^*(y_i,f(x_i),r_i)\right\}.
\end{equation}

\section{Implementation Details}\label{sec: id}
The use \eqref{eq: ipw} or \eqref{eq: BJ} for the implementation of \eqref{eq: erfal} requires consistent estimation of $\pi(y,x)$ and $f(y|X=x, R=1)$. In this section, we describe methods for estimating them and present a boosting prediction procedure for handling MNAR data.

\subsection{Construction of Consistent Estimators}
\label{subsec: cce}
We estimate $\pi(y,x)$ by modeling it parametrically. Suppose $\pi(y,x)$ is correctly modeled by a parametric form, say $\pi(y,x;\gamma)$ with the parameter $\gamma$, as stated in identification condition (A1) in the Supplementary Materials. The likelihood of $\gamma$ derived from the binary missing data indicator is 
\begin{equation*}
L(\gamma)\triangleq\prod^n_{i=1}\left\{\pi(y_i,x_i;\gamma) \right\}^{r_i}\left\{1-\pi(y_i,x_i;\gamma)\right\} ^{1-r_i},
\end{equation*}
yielding the score function
\begin{equation}
\label{eq: score}
S(\gamma)\triangleq\frac{\partial \log L( \gamma)}{\partial \gamma}=\sum^n_{i=1}S_{r_i}(y_i,x_i; \gamma), 
\end{equation} 
where
\begin{equation}
\label{eq: sr}
S_{r}(y,x;\gamma)\triangleq\left\{\frac{\partial\pi(y,x;\gamma)}{\partial\gamma}\right\}\left[\frac{r-\pi(y,x;\gamma)}{\pi(y,x;\gamma)\{1-\pi(y,x;\gamma)\}}\right]
\end{equation} 
is a vector of the same dimension as $\gamma$.

Since some $y_i$ values are missing (i.e., when $r_i=0$), directly setting \eqref{eq: score} to 0 to solve for $\gamma$ is not feasible. To address this, \cite{MorikawaKim2021} proposed an alternative estimation method. Let $S_0(Y,X; \gamma)$ denote \eqref{eq: sr} with $r$ set to 0. Let  $$O(y,x;\gamma)=\frac{1-\pi(y,x;\gamma)}{\pi(y,x;\gamma)}$$
and
\begin{equation}
\label{eq: e*}
E^*\{S_0(Y,X; \gamma)|X=x\}\triangleq\frac{E\{\pi^{-1}(Y,X;\gamma)O(Y,X;\gamma)S_0(Y,X;\gamma)|X=x,R=1\}}{E\{\pi^{-1}(Y,X;\gamma)O(Y,X;\gamma)|X=x,R=1\}}.
\end{equation}
Let $\hat{E}^*\{S_0(Y,X;\gamma)|X=x\}$ represent an estimate of \eqref{eq: e*}, with $E(\cdot|X=x, R=1)$ estimated using the conditional probability density function $f(y|X=x, R=1)$.

Solving
\begin{equation}
\label{eq: ee}
\sum^n_{i=1}\left[\left\{1-\frac{r_i}{\pi(y_i,x_i;\gamma)}\right\}\hat{E}^*\{S_0(Y_i,X_i; \gamma)|X_i=x_i\}\right]=0
\end{equation}
for $\gamma$ gives an estimator of $\gamma$. While this approach estimates $\gamma$, its implementation requires evaluation of \eqref{eq: e*}, which relies on the availability of $f(y|X=x, R=1)$. In the remainder of this subsection, we outline parametric and nonparametric estimation procedures for $f(y|X=x, R=1)$.

First, we model $f(y|X=x,R=1)$ by a parametric model, say $f(y|X=x,R=1;\beta)$ with the parameter $\beta$. The estimator of $\beta$, denoted $\hat{\beta}_{\text{p}}$, can be obtained by maximizing the conditional likelihood:
\begin{equation*}
\hat{\beta}_{\text{p}}=\argmax_{\beta}\left\{\sum^n_{i=1}r_i\log f(y_i|X_i=x_i,R_i=1;\beta)\right\}.
\end{equation*}
Thus, the estimate of $f(y|X=x,R=1)$, denoted $\hat{f}_{\text{p}}(y|X=x,R=1)$, is given by $f(y|X=x, R=1; \hat{\beta}_{\text{p}})$.

With $\hat{\beta}_{\text{p}}$, \eqref{eq: e*} can be estimated accordingly. If the conditional expectations in \eqref{eq: e*} can be derived in closed-form, then $\hat{E}^*\{S_0(Y,X; \gamma)|X=x\}$ in \eqref{eq: ee} is estimated as:
\begin{equation}
\label{eq: e*pcf}
\hat{E}_\text{p}^*\{S_0(Y,X; \gamma)|X=x\}=\frac{E\{\pi^{-1}(Y,X;\gamma)O(Y,X;\gamma)S_0(Y,X;\gamma)|X=x,R=1; \hat{\beta}_{\text{p}} \}}{E\{\pi^{-1}(Y,X;\gamma)O(Y,X;\gamma)|X=x,R=1;\hat{\beta}_{\text{p}}\}},
\end{equation}
where the conditional expectations are evaluated with respect to $f(y|X=x, R=1; \hat{\beta}_{\text{p}})$. Otherwise, $\hat{E}^*\{S_0(Y,X; \gamma)|X=x\}$ can be approximated using the fractional weights approach \citep{Kim2011} by
\begin{equation}
\label{eq: e*p}
\begin{split}
&\hat{E}_{\text{p}}^*\{S_0(Y,X; \gamma)|X=x\}\\
= \ & \sum^n_{k=1}\left[\frac{r_k\pi^{-1}(y_k,x; \gamma)O(y_k,x; \gamma)\hat{f}_{\text{p}}(y_k|X=x, R=1)/C\left(y_k,x\right)}{\sum^n_{l=1}\left\{r_l\pi^{-1}(y_l,x;\gamma)O(y_l,x;\gamma)\hat{f}_{\text{p}}(y_l|X=x,R=1)/C\left(y_l,x\right)\right\}}S_0(y_k,x;\gamma)\right],    
\end{split}
\end{equation}
where $C(y,x)=\sum^n_{t=1}r_t\hat{f}_{\text{p}}(y|X_t=x,R_t=1)$.

Under identification conditions (A1) - (A4) and regularity conditions (C1) - (C6), listed in the Supplementary Materials, a consistent estimator of $\gamma$, denoted $\hat{\gamma}_{\text{p}}$, can 
be obtained by solving \eqref{eq: ee}, with $\hat{E}^*\{S_0(Y,X; \gamma)|X=x\}$ replaced by \eqref{eq: e*pcf} or \eqref{eq: e*p}. Hence, $\hat{\pi}_{\text{p}}(y,x)\triangleq\pi(y,x;\hat{\gamma}_{\text{p}})$ provides a consistent estimate for $\pi(y,x)$. While the performance of $\hat{f}_{\text{p}}(y|X=x,R=1)$ may be sensitive to the assumed parametric form $f(y|X=x,R=1;\beta)$, \cite{MorikawaKim2021} justified that $\hat{\gamma}_{\text{p}}$ remains consistent even if $f(y|X=x,R=1;\beta)$ is misspecified. Moreover, when $f(y|X=x,R=1;\beta)$ is correctly specified,  $\hat{\gamma}_{\text{p}}$ attains the semiparametric efficiency bound. 

Alternatively, $f(y|X=x,R=1)$ can be estimated nonparametrically to avoid possible model misspecification. For illustration, consider the case where $Y_i$ is continuous and $X_i$ is univariate and continuous; the procedure can be easily generalized to discrete variables or multivariate $X_i$. Using \textit{kernel density estimation}, $f(y,x|R=1)$ and $f(x|R=1)$ can be estimated by:
\begin{equation*}
\hat{f}_{\text{np}}(y,x|R=1)=n^{-1}\sum^n_{k=1}r_kK_{h_x}(x-x_k)K_{h_y}(y-y_k),
\end{equation*}
and
\begin{equation*}
\hat{f}_{\text{np}}(x|R=1)=n^{-1}\sum^n_{l=1}r_lK_{h_x}(x-x_l),
\end{equation*}
where $K_h(u)=K\left(\frac{u}{h}\right)$, with $K(\cdot)$ denoting a kernel function and $h$ denoting the bandwidth. Here, $h_x$ and $h_y$ are bandwidths corresponding to $X_i$ and $Y_i$, respectively. Consequently, \eqref{eq: e*} can be estimated using the \textit{Nadaraya-Watson estimator}:
\begin{align}
&\ \hat{E}_{\text{np}}^*\{S_0(Y_i,X_i; \gamma)|X_i=x_i\}\notag\\
= &\ \frac{\sum^n_{k=1}r_kK_{h_x}(x_i-x_k)\pi^{-1}(y_k,x_i;\gamma)O(y_k,x_i;\gamma)S_0(y_k,x_i;\gamma)}{\sum^n_{l=1}r_lK_{h_x}(x_i-x_l)\pi^{-1}(y_l,x_i; \gamma)O(y_l,x_i;\gamma)}.\label{eq: e*np}
\end{align}
Under identification conditions (A1) - (A4) and regularity conditions (C1) - (C3) and (C7) - (C12), stated in the Supplementary Materials, a consistent estimator of $\gamma$, denoted $\hat{\gamma}_{\text{np}}$, which attains the semiparametric efficiency bound, can 
be obtained by solving \eqref{eq: ee} with $\hat{E}^*\{S_0(Y_i,X_i;\gamma)|X_i=x_i\}$ replaced by \eqref{eq: e*np}. Therefore, $\hat{\pi}_{\text{np}}(y,x)\triangleq\pi(y,x;\hat{\gamma}_{\text{np}})$ is a consistent estimator for $\pi(y,x)$. 

It is worth emphasizing that $\hat{f}_{\text{np}}(y|X=x,R=1)$ and $\hat{\pi}_{\text{np}}(y,x)$ are robust, as they do not rely on any parametric modeling assumptions for $f(y|X=x,R=1)$.
However, when $X$ is multivariate or high dimensional, these estimates become less practical due to increased computational complexity and the higher possibility of collinearity among covariates. 

\subsection{Algorithm for Boosting Prediction with MNAR Data}
\label{subsec: albemd}
With the estimates of $f(y|X=x,R=1)$ and $\pi(y, x)$, $f(y|X=x,R=0)$ can be estimated using \eqref{eq: rKY}, where $f(y|X=x,R=1)$ and $\pi(y, x)$ are replaced by their estimates. Let $\hat{f}(y|X=x,R=0)$ denote the resulting estimate. Let $\hat{L}^*(y_i, f(x_i), r_i)$ denote the adjusted loss functions in \eqref{eq: ipw} or \eqref{eq: BJ} with $\pi(y_i,x_i)$ and $f(y|X_i=x_i,R_i=1)$ replaced by their estimates described in Section \ref{subsec: cce}. For the conditional mean in \eqref{eq: BJ}, we employ an approximate method: we specify a large positive integer $N_y$ and independently take $N_y$ random draws, denoted $\left\{y_{i0}^{(1)}, \ldots, y_{i0}^{(N_y)}\right\}$, from $\hat{f}(y|X=x_i,R=0)$; then we  estimate $L_\text{BJ}(y_i, f(x_i),r_i)$ in \eqref{eq: BJ} by its empirical version:
\begin{equation}
\label{eq: bjmean}
L_\text{BJ}(y_i, f(x_i),r_i)= r_iL(y_i,f(x_i))+(1-r_i)N_y^{-1}\sum^{N_y}_{k=1}L\left(y_{i0}^{(k)},f(x_i)\right).
\end{equation} 
The goal is to find  $f\in\mathcal{F}$ by modifying \eqref{eq: erfal} to be
\begin{equation}
\label{eq: fal}
\hat{f}_n^{\text{AL}}=\argmin_{f\in\mathcal{F}}\left\{n^{-1}\sum^n_{i=1}\hat{L}^*(y_i,f(x_i),r_i)\right\},
\end{equation}
which can be implemented using Algorithm
\ref{alg}. 

\begin{algorithm}[h!]
\caption{Boosting Prediction for MNAR Data with A Modified Loss Function}\label{alg}
\begin{algorithmic}
\State Take an initial function $f^{(0)}\in\mathcal{F}$ and set $\eta$ as a small positive number;
\For{iteration $(m+1)$ with $m=0, 1, 2, \ldots$}
   \State (i) calculate $\partial \hat{L}^*\left(y_i,f^{(m)}(x_i),r_i\right)\triangleq\left.\frac{\partial \hat{L}^*\left(u, v, w\right)}{\partial v}\right|_{u=y_i, v=f^{(m)}(x_i), w=r_i}$ for $i=1, \ldots, n$;
    
    \State (ii) find $\hat{h}^{(m+1)}$ by solving
    \begin{equation*}
    \hat{h}^{(m+1)}=\argmin_{h^{(m+1)}\in\mathcal{C}}\left[n^{-1}\sum^n_{i=1}\left\{\partial \hat{L}^*\left(y_i,f^{(m)}(x_i),r_i\right)h^{(m+1)}(x_i)\right\}\right];
    \end{equation*}
    \State (iii) find $\hat{\alpha}^{(m+1)}$ by solving
    \begin{equation*}
    \hat{\alpha}^{(m+1)}=\argmin_{\alpha^{(m+1)}\in\mathbb{R}}\left\{n^{-1}\sum^n_{i=1}\hat{L}^*\left(y_i, f^{(m)}(x_i)+\alpha^{(m+1)}\hat{h}^{(m+1)}(x_i),r_i\right)\right\};
    \end{equation*}
    \State (iv) update 
    $f^{(m+1)}(x_i)=f^{(m)}(x_i)+\hat{\alpha}^{(m+1)}\hat{h}^{(m+1)}(x_i)$ for $i=1, \ldots, n$;
    \NoThen
    \If{at iteration $\tilde{m}+1$},
    \begin{equation}
    \label{eq: stopal}
    \left|n^{-1}\sum^n_{i=1}\hat{L}^*\left(y_i,f^{(\tilde{m})}(x_i),r_i\right)-n^{-1}\sum^n_{i=1}\hat{L}^*\left(y_i, f^{(\tilde{m}+1)}(x_i),r_i\right)\right|\leq \eta
    \end{equation}
    \State \textbf{then} stop iteration and define the final estimator as
    \begin{equation*}
    \hat{f}_n^{\text{AL}}(\cdot)=f^{(\tilde{m})}(\cdot)
    \end{equation*}
\EndIf
\EndFor
\end{algorithmic}
\end{algorithm}

\subsection{Specification of the class \texorpdfstring{$\mathcal{C}$}{C}}\label{subsec: sc}
The boosting prediction procedure for MNAR data described in Section \ref{subsec: albemd} depends on the specification of the class 
$\mathcal{C}$. Similar to \cite{ChenYi2024}, we consider the regression spline method to characterize the functions in $\mathcal{C}$. Instead of using the truncated power basis functions, which are known to be numerically unstable in practice \citep{HastieTibshirani2009}, we employ more computationally stable B-spline basis functions \citep{deboor2001}. 

Specifically, any function $h(\cdot)$ in $\mathcal{C}$ is expressed in an additive form:
\begin{equation*}
h(x)=h_0+h_1(x_{1})+\ldots+h_p(x_{p}),
\end{equation*}
with $h_0$ denoting the intercept and
$h_k(x_{k})=\sum^{T}_{t=1}\xi_{kt}b_{kt}(x_{k})$ for $k=1, \ldots, p$, where $x=(x_{1}, \ldots, x_{p})^\top$; $b_k(x_{k})=(b_{k1}(x_{k}), \ldots, $ $b_{kT}(x_{k}))^\top$ is the vector of the B-spline basis functions of order $M$, with $T-M+1$ interior knots at suitably chosen quantiles of $X_{k}$ and $M\geq2$ being an integer; and $\xi_k=(\xi_{k1}, \ldots, \xi_{kT})^\top$ is the unknown parameter vector. 

\section{Theoretical Results}\label{sec: tr}
In this section, we develop theoretical results for the proposed methods, based on the assumption that $\pi(y,x)$ and $f(y|X=x, R=1)$ are consistently estimated, as stated in condition (B5) in the Supplementary Materials. First, we discuss the convergence of the proposed iterated algorithm, described in Algorithm \ref{alg}, and defer the proofs to the Supplementary Materials.

\begin{proposition}\label{prop2}
Assume regularity conditions (B1) - (B8) in the Supplementary Materials. Suppose that Algorithm \ref{alg} is run to a random sample $\mathcal{O}_\text{missing}$ with the given size $n$
considered in Section \ref{sec: ubemnar}. For any initial function $f^{(0)}\in\mathcal{F}$, let $f^{(m+1)}$ denote the updated estimate of the function at iteration $(m+1)$ of Algorithm \ref{alg}. Then there exist positive constants $c^*$ and $C^*$ with $c^*C^*>1$ such that 
\begin{equation}
\label{eq: ralbdd}
R\left(f^{(m+1)}\right)-R\left(f^*\right)\leq\left(1-\frac{1}{C^*c^*}\right)^m\left\{R\left(f^{(0)}\right)-R(f^*)\right\}, 
\end{equation} 
where $R(\cdot)$ and $f^*$ are defined in \eqref{eq: rf} and \eqref{eq: rfmin}, respectively. 
\end{proposition}

Proposition \ref{prop2} demonstrates that for Algorithm \ref{alg} and any $m=0,1,2,\ldots$, the difference between $R\left(f^{(m+1)}\right)$ and $R\left(f^*\right)$ is upper bounded by the difference between $R\left(f^{(0)}\right)$ and $R\left(f^*\right)$ that is multiplied by the power $m$ of a positive constant smaller than 1. As $m\to\infty$, this upper bound approaches zero, as summarized as follows.

\begin{theorem}\label{thm1}
Under the setup of Proposition \ref{prop2},
\begin{equation*}
\lim_{m\to\infty}R\left(f^{(m+1)}\right)=R\left(f^*\right).
\end{equation*}
\end{theorem}

While the convergence of Algorithm \ref{alg} is guaranteed by Theorem \ref{thm1},  in applications, Algorithm \ref{alg} stops at a finite number to avoid excessively running an unnecessarily large number of iterations. Similar to \cite{ChenYi2024}, we may use \eqref{eq: ralbdd}, in combination with \eqref{eq: stopal}, to decide an upper bound for the stopping iteration number $\tilde{m}$ in Algorithm \ref{alg}:
\begin{equation}
\label{eq: ub}
\tilde{m}<1+\frac{\log\left\{\frac{\eta}{\left|R\left(f^{(0)}\right)-R(f^*)\right|\left(2-\frac{1}{C^*c^*}\right)}\right\}}{\log\left(1-\frac{1}{C^*c^*}\right)},
\end{equation}
where $\eta$ is a given threshold in Algorithm \ref{alg}. The upper bound of \eqref{eq: ub} shows that the number of iterations may vary with the choice of an initial function $f^{(0)}$ and the required accuracy $\eta$, as noted by \cite{BoydVandenberghe2004} and \cite{ChenYi2024}.

Next, we show the consistency of the estimator $\hat{f}_n^\text{AL}$, defined in \eqref{eq: fal}.

\begin{theorem}\label{thm2}
Assume the conditions in Theorem \ref{thm1} and condition (B9) in the Supplementary Materials. Suppose that Algorithm \ref{alg} is run to a sequence of random samples with a varying size $n$. Then for any $\epsilon>0$,
\begin{equation*}
P\left(\left\|\hat{f}_n^\text{AL}-f^*\right\|_{\infty}\leq\epsilon\right)\xrightarrow[\ \ ]{}1 \text{ as } n\to\infty,
\end{equation*}
where $\left\|\hat{f}_n^\text{AL}-f^*\right\|_{\infty} = \max_{1\leq i\leq n}\left|\hat{f}_n^\text{AL}(X_i)-f^*(X_i)\right|$ is the $L_\infty$ norm of $\hat{f}_n^\text{AL}-f^*$, evaluated over $\{X_i: i=1, \ldots, n\}$.
\end{theorem}

Theorem \ref{thm2} indicates that the difference between the proposed estimator $\hat{f}_n^\text{AL}$ and its target $f^*$, expressed in terms of the $L_\infty$-norm, converges in probability as $n\to\infty$. 

\section{Simulation Studies} \label{sec: ss}

In this section, we assess the finite sample performance of the proposed methods. Five hundreds simulations are run for each parameter configuration discussed below, with the sample size $n$ set to $1000$.

\subsection{Simulation Design and Data Generation}\label{subsec: sddg}
We consider two settings, as in \cite{BianYi2025a}.  with different dimensions $p$ for covariates $X_i$: Setting 1 has $p=2$, and Setting 2 includes a larger number of covariates with $p=9$.

For the two settings, we use slightly different ways to generate covariates $X_i$ yet the same procedure to generate responses $Y_i$ and missing indicators $R_i$. In Setting 1, we generate $X_{1, i}$ independently from $N\left(0, \sigma_X^2\right)$ for $i=1, \ldots, n$; and given $X_{1, i}$, $X_{2, i}$ is drawn from $N\left(\zeta X_{1, i}, \sigma_X^2\right)$, with $\zeta=-0.25$ and $\sigma_X^2=\frac{1}{2}$. In Setting 2, $X_i=(X_{1,i},\ldots,X_{9,i})^\top$ is independently generated for $i=1,\ldots, n$ from the multivariate normal distribution with zero mean and covariance matrix given by the 9 × 9 matrix with element $(u, v)$ being $0.5^{|u-v|}/\sqrt{2}$ for $u,v=1,\ldots,9$.

For Setting $j$ with $j=1,2$, given the covariates $X_i$, we generate the missing data indicator $R_i$ from Bernoulli($\tilde{\pi}_j(X_i)$), with  
\begin{align}
\tilde{\pi}_j(X_i)= \frac{1}{1+\exp\{g_j(X_i)\}}\quad
\text{and}\quad
g_j(X_i)= \frac{1}{2}\gamma_y^2\sigma^2-\nu_j(X_i)-\gamma_y\mu_j(X_i),\label{eq: gx}
\end{align}
where for Setting 1, we specify $\nu_1(X_i)=\gamma_{1,0}+\gamma_{1,1}X_{1, i}+\gamma_{1,2}X_{2, i}$ and $\mu_1(X_i)=\beta_{1,0}+\beta_{1,1}X_{1, i}+\beta_{1,2}X_{2, i}+\beta_{1,3}X_{1, i}X_{2, i}$ with $\beta_{1,1}=\beta_{1,2}=0.4$ and $\gamma_{1,1}=\gamma_{1,2}=-0.5$; and for Setting 2, we set $\nu_2(X_i)=\gamma_{2,0}+\sum^8_{k=1}\gamma_{2,k}X_{k, i}$ and $\mu_2(X_i)=\beta_{2,0}+\sum^9_{k=1}\beta_{2,k}X_{k, i}$ with $\beta_{2,1}=\ldots=\beta_{2,9}=0.4$ and $\gamma_{2,1}=\ldots=\gamma_{2,8}=-0.5$. For both settings, $\sigma=0.5$.

Condition (A4) in the Supplementary Materials implies that setting $\beta_{1,3}\neq0$ ensures the model to be identifiable. Noted by \cite{MorikawaKim2021} through simulation studies, almost half of the simulations failed to converge if $\beta_{1,3}$ is taken as 0. Consequently, here we only consider settings with a nonzero $\beta_{1,3}$, set as $\beta_{1,3}=0.8$. 

We set $\gamma_{1,0}=\gamma_{2,0}=0.405$ to achieve approximately 40\% missing observations at the baseline when taking $X_{i}$ and $Y_i$ to be zero in both settings. Furthermore, to represent different missing data mechanisms, for Setting 1, we set $\gamma_y=0$ to represent an MAR and set $\gamma_y=-1$ to reflect an MNAR; and for Setting 2, we set $\gamma_y=0$ for an MAR and $\gamma_y=1$ for an MNAR. The average missingness proportions in our simulated samples are about 40.4\%, 41.3\%, 43.6\%, and 41.1\% for Setting 1 (MAR), Setting 1 (MNAR), Setting 2 (MAR), and Setting 2 (MNAR), respectively.

As shown in the Supplementary Materials, in Setting 1, by taking $E(Y_i)=0$, $\beta_{1,0}$ is determined by
\begin{equation}
\label{eq: beta0}
\gamma_y\sigma^2\{1-\Pr(R_i=1)\}-\zeta \beta_{1,3}\sigma^2_X,
\end{equation}
leading to $\beta_{1,0}=0.1$ for the MAR scenario and $\beta_{1,0}=0$ for the MNAR scenario, if we set $\Pr(R_i=1)=0.6$. Similarly, for Setting 2, if $E(Y_i)=0$ and $\Pr(R_i=1)=0.6$, $\beta_{2,0}$ is determined by  
\begin{equation}
\label{eq: beta0s2}
\gamma_y\sigma^2\{1-\Pr(R_i=1)\},
\end{equation}
as detailed in the Supplementary Materials. Thus, $\beta_{2,0}=0$ for the MAR scenario and $\beta_{2,0}=0.1$ for the MNAR case.

Consequently, given $X_i$ and $R_i$ for $i=1,\ldots,n$, the response $Y_i$ is generated from 
\begin{align}
\label{eq: y}
N(\mu_j(X_i)-(1-R_i)\gamma_y\sigma^2,\sigma^2),
\end{align} 
as detailed in the Supplementary Materials.

Next, in both settings, the generated data are split randomly by the $4:1$ ratio, and we let $\mathcal{O}^\text{TR}\triangleq\{y_i,x_i,r_i:i=1,\cdots,n_1\}$ and $\mathcal{O}^\text{TE}\triangleq\{y_i, x_i, r_i: i=n_1+1,\cdots,n_1+n_2\}$ denote them, respectively, where $n_1=800$ and $n_2=200$. The training data $\mathcal{O}^\text{TR}_\text{missing}$, which includes $\{y_i,x_i,r_i=1\}$ or $\{x_i,r_i=0\}$ with $i=1, \ldots, n_1$, are used to obtain the estimated $\hat{f}_{n_1}^*(\cdot)$; and the test data $\mathcal{O}^\text{TE}_\text{missing}$, which includes $\{y_i,x_i,r_i=1\}$ or $\{x_i,r_i=0\}$ with $i=n_1+1, \ldots, n_1+n_2$, are utilized to evaluate the performance of $\hat{f}_{n_1}^\text{PO}(\cdot)$. Define $\mathcal{O}^\text{TR}_\text{full}\triangleq \{y_i,x_i: i=1,\cdots,n_1\}$.

\subsection{Evaluation Metrics}\label{subsec: em}
To evaluate the performance of the proposed boosting prediction methods, we calculate the following two metrics for the target function $f^*$ as in \eqref{eq: rfmin}.

The first metric represents the sample-based maximum absolute error (S-MAE) \citep{ChenYi2024}, defined as the infinity norm of the difference between the estimate and the true function with respect to the sample:
\begin{equation*}
\left\|\hat{f}_{n_1}^*-f^*\right\|_{\infty} = \max_{n_1+1\leq i\leq n_1+n_2}\left|\hat{f}_{n_1}^*(x_i)-f^*(x_i)\right|,
\end{equation*} 
and the second metric reports square root of the sample-based mean squared error (S-RMSE), defined as:
\begin{equation*}
\left\|\hat{f}_{n_1}^*-f^*\right\|_2 = \sqrt{{n_2}^{-1}\sum^{n_1+n_2}_{i=n_1+1}\left\{\hat{f}_{n_1}^*(x_i)-f^*(x_i)\right\}^2}.   
\end{equation*} 
These two metrics take different perspectives to reflect the discrepancies of $\hat{f}^*_{n_1}$ from its target $f^*$.

\subsection{Analysis Methods}\label{subsec: am}
We analyze the simulated data using five methods, where we set $M=2$ and $T=3$ for the specification of the class $\mathcal{C}$, discussed in Section \ref{subsec: sc}. For the stopping criterion, we set $\eta=10^{-6}$. 

The first two methods, called \textit{R} (reference) and \textit{N} (naive), address two extreme scenarios: applying the conventional boosting procedure introduced in Section \ref{subsec: cbep} to the full dataset $\mathcal{O}_\text{full}^\text{TR}$ and to the dataset $\mathcal{O}_\text{missing}^\text{TR}$, respectively. The next three methods apply boosting with modified loss functions described in Algorithm \ref{alg}. Specifically, we employ the inverse propensity weight adjusted loss functions \eqref{eq: ipw} with parametrically modeled $f(y|X=x,R=1)$ and nonparametrically estimated $f(y|X=x,R=1)$, respectively, denoted \textit{IPW} and \textit{IPWN}, where nonparametric estimation uses a Gaussian kernel with a rule-of-thumb bandwidth $h_x=(\sum_{i=1}^{n_1}r_i)^{-\frac{1}{5}}\hat{\sigma}_x$, with $\hat{\sigma}_x$ denoting the sample standard deviation of the covariate, as suggested by \cite{MorikawaKim2021}. Additionally, we consider the Buckley-James-type adjusted loss function \eqref{eq: BJ}, referred to as the \textit{BJ} method, employing \eqref{eq: bjmean} with $N_y=20$.

Let L1, L2, and H denote the $L_1$ loss, the $L_2$ loss, and the Huber loss, respectively, given by
$L\left(Y_i, f(X_i)\right)=\left|Y_i-f(X_i)\right|$, $L\left(Y_i, f(X_i)\right)=\frac{1}{2}\left\{Y_i-f(X_i)\right\}^2$,
and 
\begin{equation*}
L\left(Y_i, f(X_i)\right)=\begin{cases} 
      \frac{1}{2}\left\{Y_i-f(X_i)\right\}^2, & \text{if } \left|Y_i-f(X_i)\right|\leq\eta, \\
     \eta\left(\left|Y_i-f(X_i)\right|-\frac{\eta}{2}\right), & \text{otherwise},
   \end{cases}
\end{equation*}
where $\eta$ is a transition point, which can be chosen as the $\alpha$th quantile of the set $\{|Y_i-f(X_i)|: i=1,\ldots,n\}$, with $0\leq\alpha\leq100$ \citep{Friedman2001}. For the Huber loss,  $\eta$ is iteratively specified as the 50th-quantile of $\{R_i|Y_i-f^{(m-1)}(X_i)|: i=1,\ldots,n\}$ at iteration $m$, as in \cite{Friedman2001} and \cite{BuhlmannHothorn2007}. As discussed by \cite{BuhlmannHothorn2007} and \cite{HastieTibshirani2009}, the solution $f^{*}$ in \eqref{eq: rfmin} corresponds to median$(Y_i|X_i)$ and mean$(Y_i|X_i)$ for the $L_1$ loss and $L_2$ loss functions, respectively. Here ``median$(Y_i|X_i)$'' and ``mean$(Y_i|X_i)$'' represent the median and mean of the conditional probability function, $f_j(y|X_i)$, of $Y_i$ given $X_i$, which is given by
\begin{align}
f_j(y|X_i)= \tilde{\pi}_j(X_i)f_j(y|X_i,R_i=1)+\{1-\tilde{\pi}_j(X_i)\}f_j(y|X_i,R_i=0)\label{eq: fyx}
\end{align}
for Setting $j=1$ or 2, where $f_j(y|X_i,R_i=0)$ and $f_j(y|X_i,R_i=1)$ is defined in the Supplementary Materials. As shown in the Supplementary Materials, corresponding to Setting $j$ with $j=1,2$, mean$(Y_i|X_i)$ equals 
\begin{align}
\mu_j(X_ i)-\{1-\tilde{\pi}_j(X_i)\}\gamma_y\sigma^2,\label{eq: fminl2}
\end{align}
and median$(Y_i|X_i)$ can be obtained by solving 
$\int^q_{-\infty}f_j(y|X_i)dy=0.5$
for $q$.

In the MAR scenario, since $f_j(y|X_i)=f_j(y|X_i,R_i=1)=f_j(y|X_i,R_i=0)$, which is identical to the probability density function (pdf) of (S.23) in the Supplementary Materials, median$(Y_i|X_i)$
coincides with mean$(Y_i|X_i)$ in \eqref{eq: fminl2}. Given that 
the Huber loss behaves as the $L_2$
loss when $\left|Y_i-f(X_i)\right|\leq\eta$, and as a linear form of the $L_1$ loss otherwise, the solution $f^*$ in \eqref{eq: rfmin} also corresponds to \eqref{eq: fminl2} if the Huber loss is used. However, in the MNAR scenario, median$(Y_i|X_i)$ does not necessarily equal mean$(Y_i|X_i)$, so $f^*$ generally lacks an analytic form when using the Huber loss.

\subsection{Simulation Results}\label{subsec: sr}
\begin{figure}[h!]
    \centering
    \includegraphics[width=\textwidth]{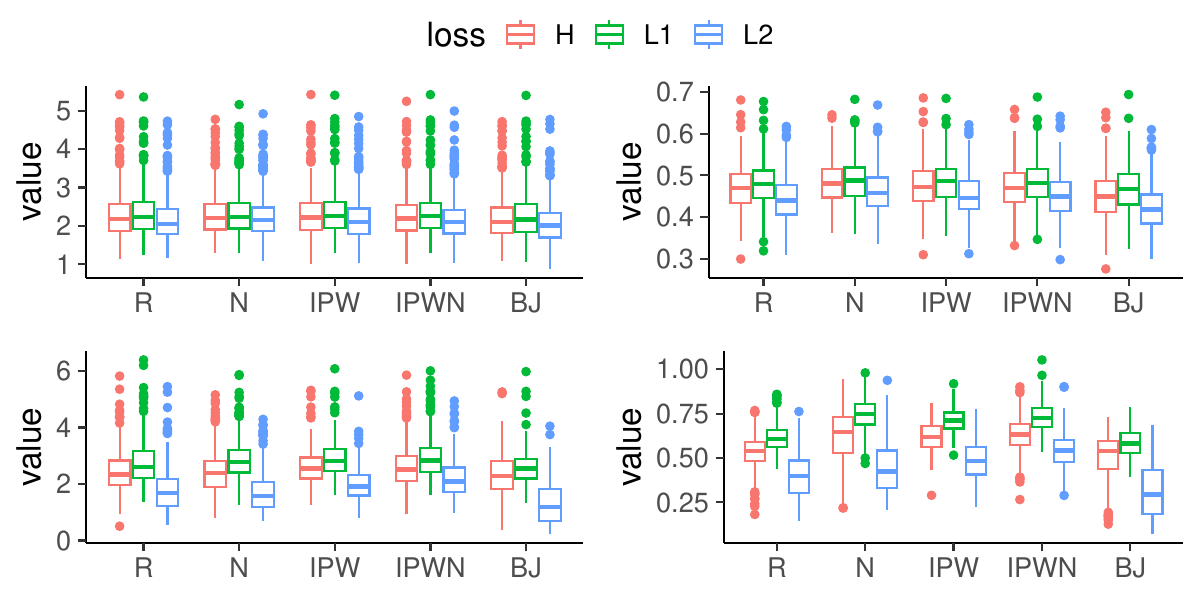}\vspace{-9mm}
    \caption{{\it Prediction assessments in the MAR scenario. Top and bottom rows correspond to Settings 1 and 2, respectively; left and right columns correspond to the values for S-MAE and S-RMSE, respectively.}}
    \label{fig: metricsMAR}
\end{figure}

\begin{figure}[h!]
    \centering
    \includegraphics[width=\textwidth]{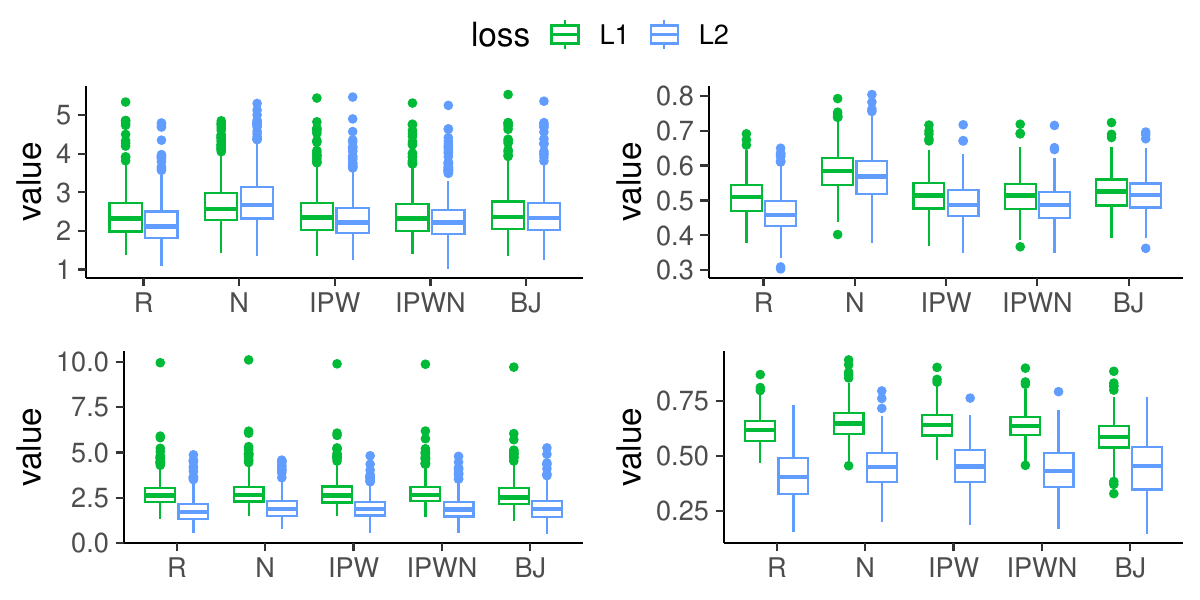}\vspace{-9mm}
    \caption{\it Prediction assessments in the MNAR scenario. Top and bottom rows correspond to Settings 1 and 2, respectively; left and right columns correspond to the values for S-MAE and S-RMSE, respectively.}
    \label{fig: metrics}
\end{figure}

In Figures \ref{fig: metricsMAR} and \ref{fig: metrics}, we summarize the S-MAE and S-RMSE values for 500 simulations as boxplots. Figure \ref{fig: metricsMAR} presents results for the MAR scenario with three loss functions, while Figure \ref{fig: metrics} focuses on the MNAR scenario using only $L_1$ and $L_2$ loss functions. In Figure \ref{fig: metricsMAR} with MAR, all methods perform similarly, with the $L_2$ loss yielding the smallest median of S-MAE and S-RMSE values. In Figure \ref{fig: metrics} with MNAR, for Setting 1, the N methods produce the largest S-MAE and S-RMSE, whereas for Setting 2, all methods using the same loss function yield similar S-MAE and S-RMSE values. The $L_1$ loss function produces larger values and the $L_2$ loss function yields smaller values. The IPW, IPWN, and BJ methods provide comparable results to the R methods. 
\vspace{-5mm}

\begin{figure}[h!]
    \centering
    \includegraphics[width=\textwidth]{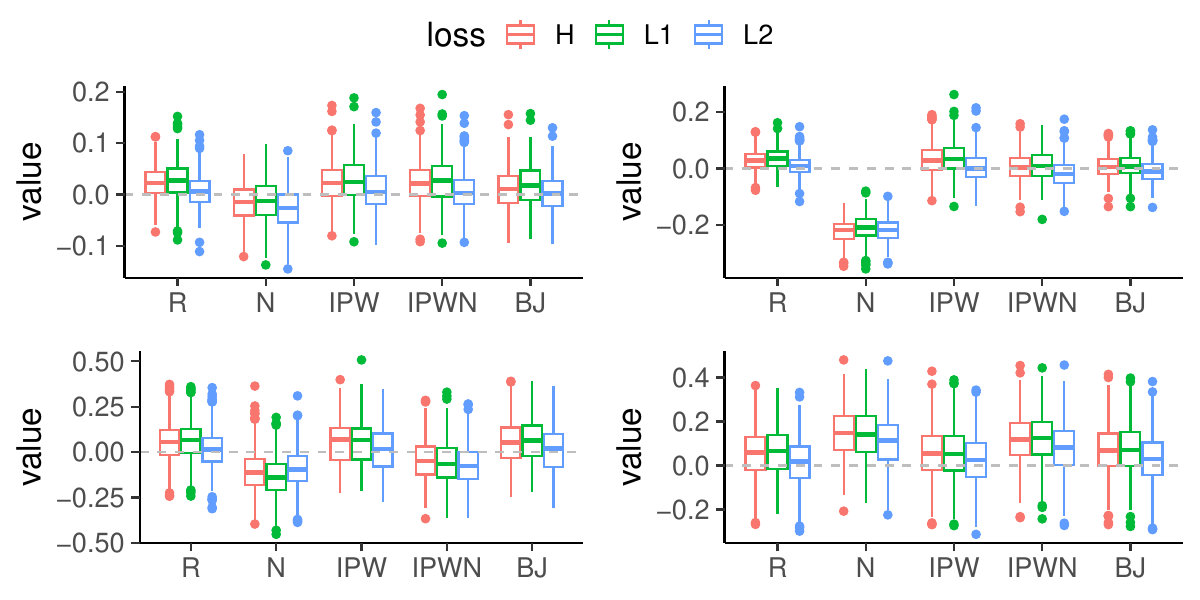}\vspace{-9mm}
    \caption{\it Boxplots of ${n_2}^{-1}\sum^{n_2}_{i=1}\hat{f}^*_{n_1}(X_i)$ obtained from the five methods in combination with the three loss functions, where the dashed line indicates the value of $E(Y_i)$. Top and bottom rows correspond to Settings 1 and 2, respectively; left and right columns correspond to the MAR and MNAR settings, respectively.}
    \label{fig: smpe}
\end{figure}

To visualize predicted values derived from different methods, Figure \ref{fig: smpe} shows boxplots of the average predicted values, ${n_2}^{-1}\sum^{n_2}_{i=1}\hat{f}^*_{n_1}(X_i)$, across 500 simulations, along with the expected value $E(Y_i)$, whose expression is  provided in the Supplementary Materials. Under MAR, all methods yield similar results, though the N method tends to deviate more from the R method than the proposed methods. However, with MNAR in Setting 1, the R and proposed methods (i.e., IPW, IPWN, and BJ) produce fairly close results, while the N methods exhibit noticeable bias. In Setting 2, with a large number of correlated covariates, IPWN does not perform satisfactorily, as the MNAR estimation procedure of \cite{MorikawaKim2021}   assumes independent covariates.

Finally, we assess the convergence of the proposed boosting methods. The results confirm the convergence of our boosting algorithm, as long as estimates of $\pi_2(y,x)$, defined in the Supplementary Materials, using the method of \cite{MorikawaKim2021} are available; further details are provided in the Supplementary Materials. We note that the method proposed by \cite{MorikawaKim2021} is specifically designed for MNAR data and may not perform well for MAR data, particularly in scenarios with high-dimensional covariates. In such cases, divergence can occur frequently, as acknowledged by \cite{MorikawaKim2021}, thus preventing the estimation of $\pi_2(y,x)$ and halting the boosting procedure. This issue may stem from various factors related to starting values for the nonlinear optimization function, the complexity of the optimization process, and the large number of parameters. 

\subsection{Model Misspecification}
\label{subsec: mm}
In addition to Sections \ref{subsec: alfmda} and \ref{subsec: cce}, we further conduct a simulation study to assess the performance of the IPW and IPWN methods under model misspecification for $\pi_j(Y_i,X_i)$ and $f_j(y|X_i,R_i=1)$, with $j=1,2$. 

\begin{figure}[h!]
    \centering
    \includegraphics[width=\textwidth]{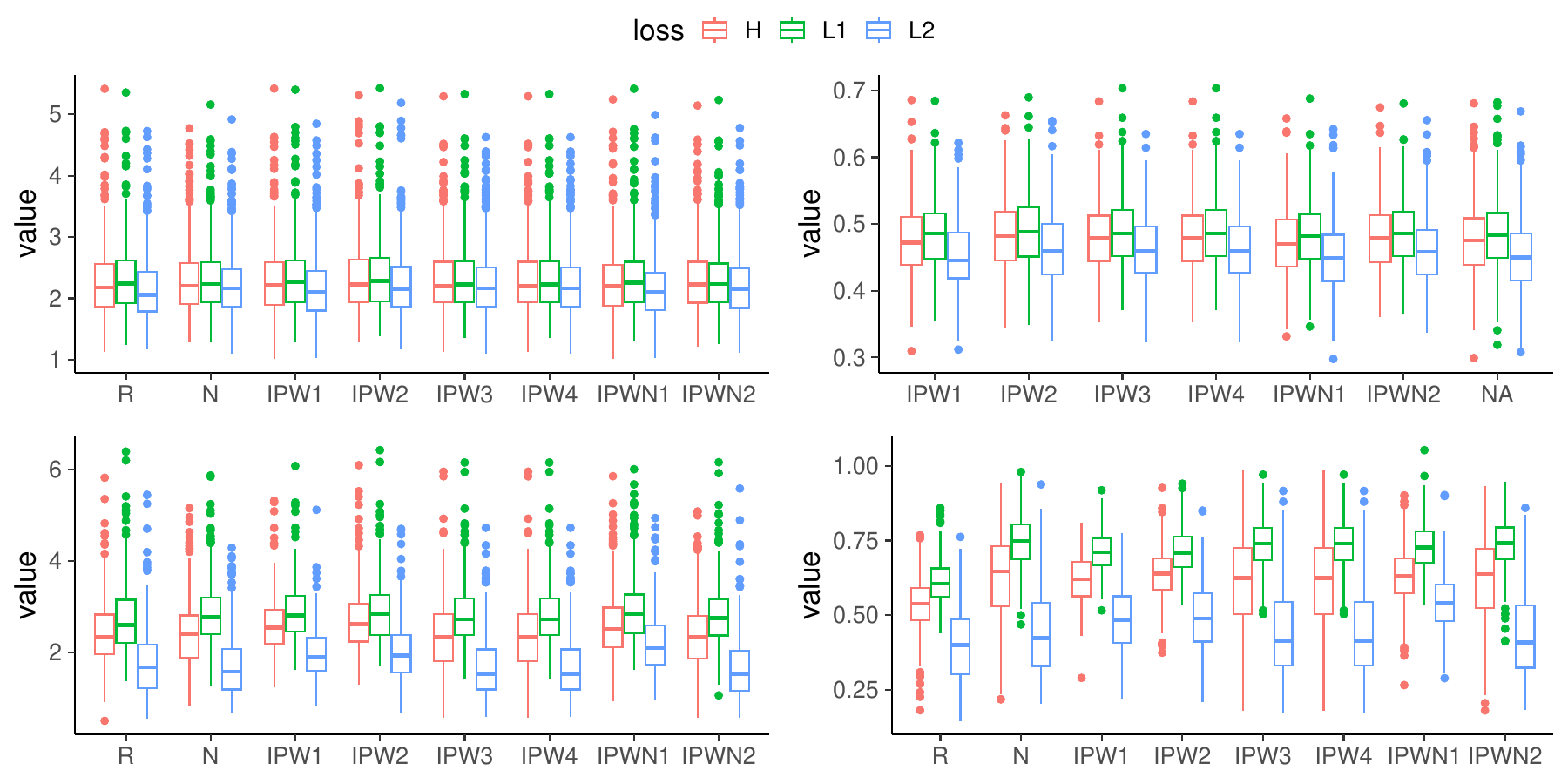}\vspace{-9mm}
    \caption{\it Performance under model misspecification in the MAR scenario. Top and bottom rows correspond to Settings 1 and 2, respectively; left and right columns correspond to the values for S-MAE and S-RMSE, respectively.}
    \label{fig: metricsMARipw}
\end{figure}

\begin{figure}[h!]
    \centering
    \includegraphics[width=\textwidth]{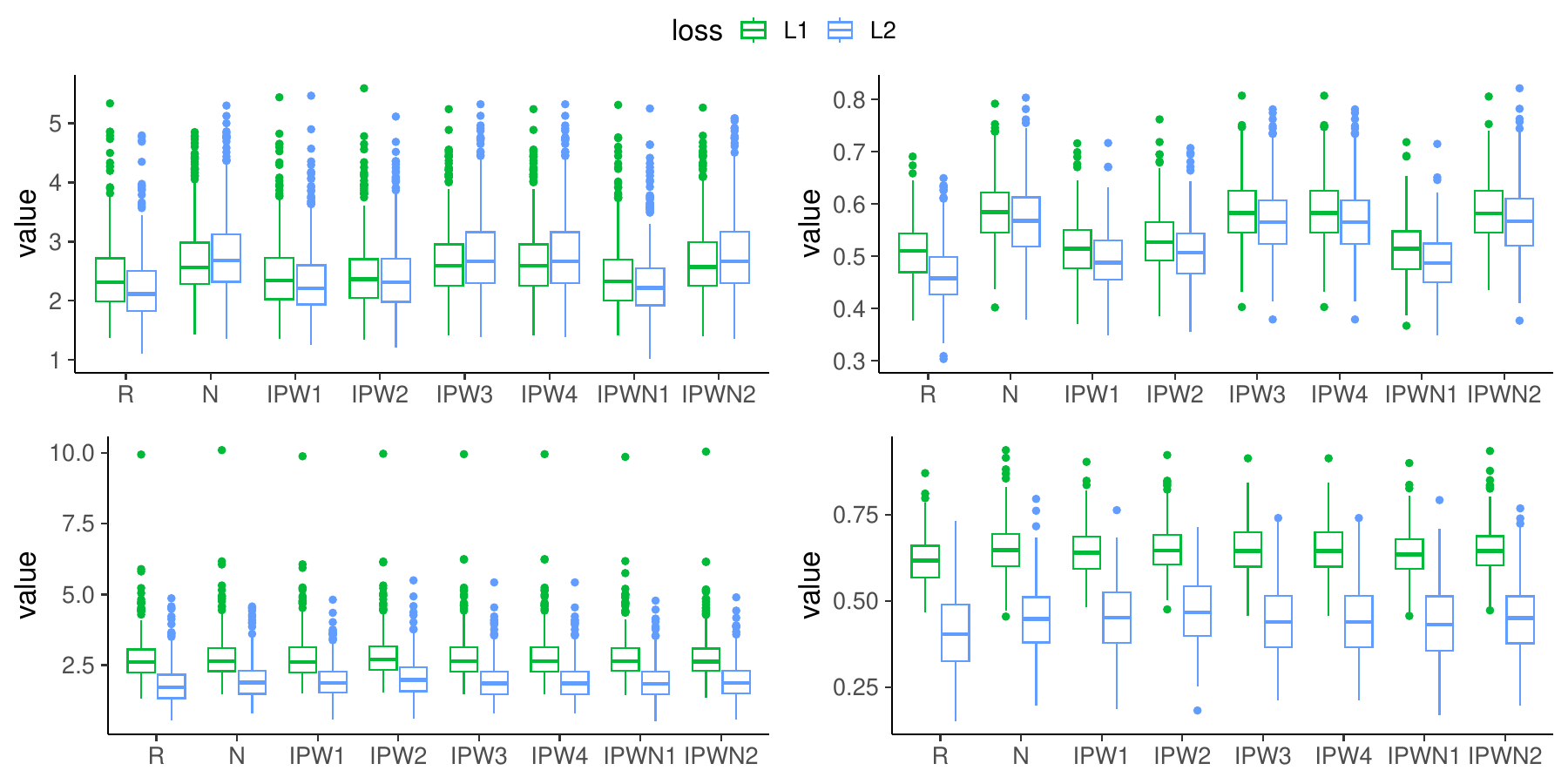}\vspace{-9mm}
    \caption{\it Performance under model misspecification in the MNAR scenario. Top and bottom rows correspond to Settings 1 and 2, respectively; left and right columns correspond to the values for S-MAE and S-RMSE, respectively.}
    \label{fig: metricsipw}
\end{figure}

\begin{figure}[h!]
    \centering
    \includegraphics[width=\textwidth]{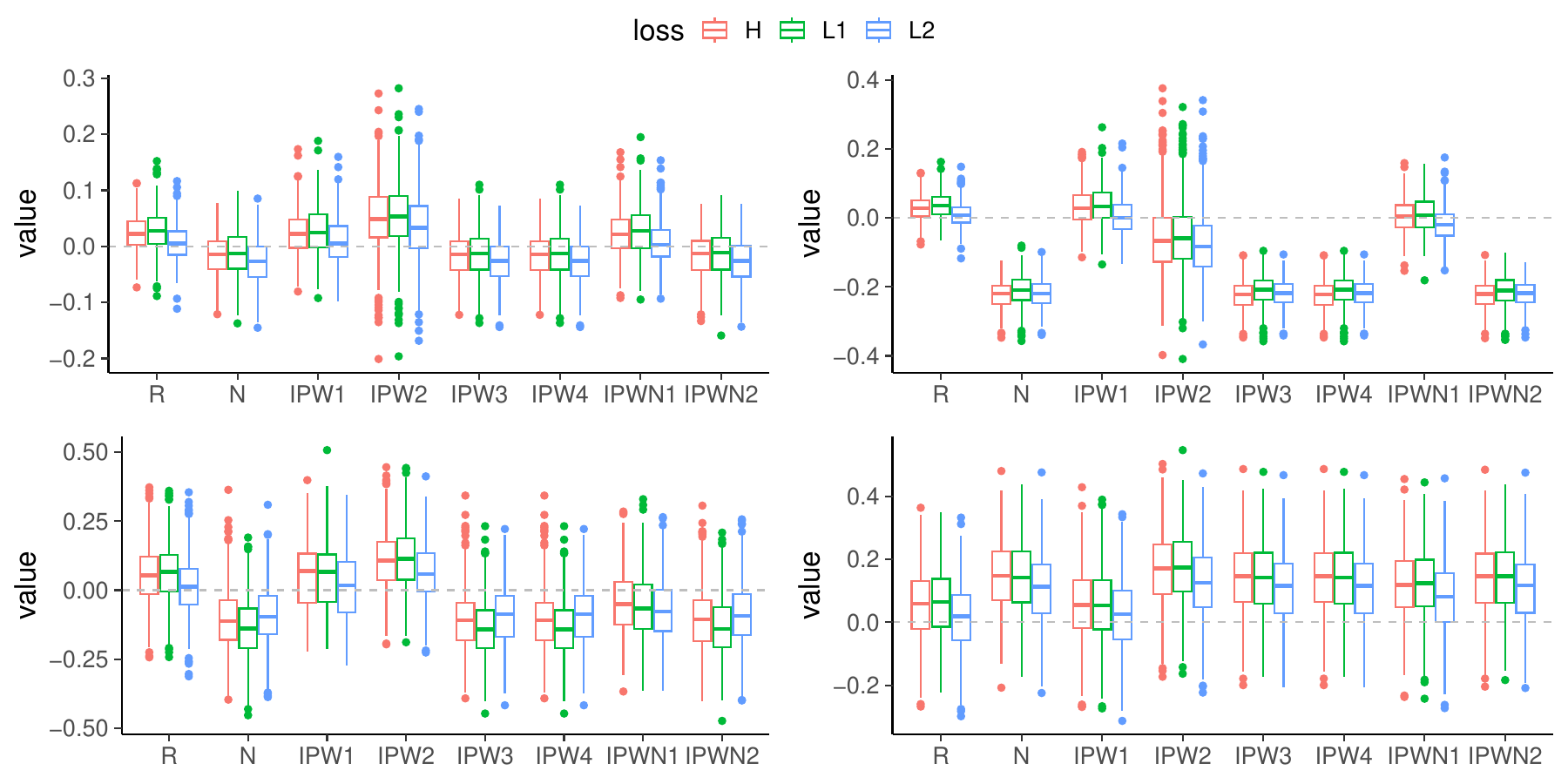}\vspace{-9mm}
    \caption{\it Performance under model misspecification: Boxplots of $n^{-1}\sum^{n}_{i=1}\hat{f}^*_n(X_i)$ obtained from the eight methods in combination with the three loss functions, where the dashed line indicates the value of $E(Y_i)$. Top and bottom rows correspond to Settings 1 and 2, respectively; left and right columns correspond to the MAR and MNAR settings, respectively.}
    \label{fig: smpeIPW}
\end{figure}

We use the same procedure as in Section \ref{subsec: sddg} to generate data, where $f_j(y|X_i,R_i=1)$ and $\pi_j(Y_i,X_i)$ are respectively taken as the pdf of (S.23) and (S.24). However, when fitting the data with the IPW methods, we intentionally misspecify $f_j(y|X_i,R_i=1)$ and $\pi_j(Y_i,X_i)$ as the pdf of $N\left(\beta_{j,1}X_{1,i},\sigma^2\right)$ and $\exp\left(\gamma_{j,0}\right)/\{1+\exp\left(\gamma_{j,0}\right)\}$, respectively. Specifically, we let IPW1, IPW2, IPW3, and IPW4 represent  cases where both $f_j(y|X_i,R_i=1)$ and $\pi_j(Y_i,X_i)$ are correctly specified, only $f_j(y|X_i,R_i=1)$ is misspecified, only $\pi_j(Y_i,X_i)$ is misspecified, and both $f_j(y|X_i,R_i=1)$ and $\pi_j(Y_i,X_i)$ are misspecified, respectively. For the IPWN method, we let IPWN1 and IPWN2 represent cases where $\pi_j(Y_i,X_i)$ is correctly specified and misspecified, respectively. The results are displayed in Figures \ref{fig: metricsMARipw} - \ref{fig: smpeIPW} in a manner similar to those in Figures \ref{fig: metricsMAR} - \ref{fig: smpe}, where in contrast, we also include results obtained from the R and N methods. 

In the MAR scenario of Setting 1, all IPW and IPWN methods perform similarly. In the MNAR scenario of Setting 1, the IPW and IPWN methods with misspecified $\pi_j\left(Y_i,X_i\right)$, namely, IPW3, IPW4, and IPWN2, exhibit performance similar to the N method, whereas the IPW and IPWN methods with correctly specified $\pi_j\left(Y_i,X_i\right)$, namely, IPW1, IPW2, and IPWN1, yield results closer to those obtained from the R method. For the IPW methods, when there is a single model misspecification (either for $\pi_j\left(Y_i,X_i\right)$ or $f_j\left(y|X_i, R_i=1\right)$), misspecifying the latter has less noticeable effects compared to misspecifying the former, which aligns with the discussions in Sections \ref{subsec: alfmda} and \ref{subsec: cce}. For the MAR scenario of Setting 2, the results are similar to those in Setting 1, with the N, IPW2, IPW3, and IPW4 methods (but not the IPWN methods) exhibiting more biased results. However, in the MNAR scenario of Setting 2, only the IPW1 method yields results comparable to the R method.

\vspace{-8mm}

\section{Analysis of KLIPS Data}\label{sec: da}
To illustrate the proposed methods, we apply them to analyze a dataset from the Korean Labor and Income Panel Survey (KLIPS), which contains demographics information for 2501 Korean workers. The KLIPS data are frequently analyzed in the MNAR literature, including \cite{KimYu2011}, \cite{WangShao2014}, \cite{ShaoWang2016}, \cite{MorikawaKim2017}, \cite{MorikawaKim2021}, and \cite{BianYi2025a}.

For subject $i$, let the response variable $Y_i$ denote the income (in $10^6$ Korean Won) of worker $i$ in 2008; let $X_{i1}$ denote the income (in $10^6$ Korean Won) of worker $i$ in 2007; let $X_{i2}$ be 1 if the age is less than 35, 2 if the age is between 35 and 51, and 3 otherwise; and let $X_{i3}$ be 1 if male and 2 if female. The response variable has about 30.63\% missing values while all covariate values are observed. 

Assume that the missing data indicator $R_i$ follows the Bernoulli distribution, Bernoulli($\pi(Y_i,X_i)$), with
\begin{equation*}
\text{logit}\{\pi(Y_i,X_i)\}=\gamma_0+\gamma_1X_{1, i}+\gamma_2X_{2, i}+\gamma_3X_{3, i}+\gamma_4Y_{i},
\end{equation*}
where $\gamma_j$ is the parameter for $j=0,1,\ldots,4$. Given the missing indicator $R_i=1$ and covariates $X_{i,1}$, $X_{i,2}=j$, $X_{i3}=k$ for $j=1,2,3$ and $k=1,2$, the response $Y_i$ follows $N(\beta_{0,j,k}+\beta_{1,j,k}X_{i,1}+\beta_{2,j,k}X_{i,1}^2+\beta_{3,j,k}X_{i,1}^3+\beta_{4,j,k}X_{i,1}^4,\sigma_{j,k}^2)$, with parameters $\beta_{j,k}=(\beta_{0,j,k}, \beta_{1,j,k}, \beta_{2,j,k}, \beta_{3,j,k}, \beta_{4,j,k})^\top$ and $\sigma_{j,k}$. Following \cite{MorikawaKim2021}, for each combination of $j$ and $k$, we use the stepwise AIC to select the best model.

\begin{figure}[h]
    \centering
    \includegraphics[width=\textwidth]{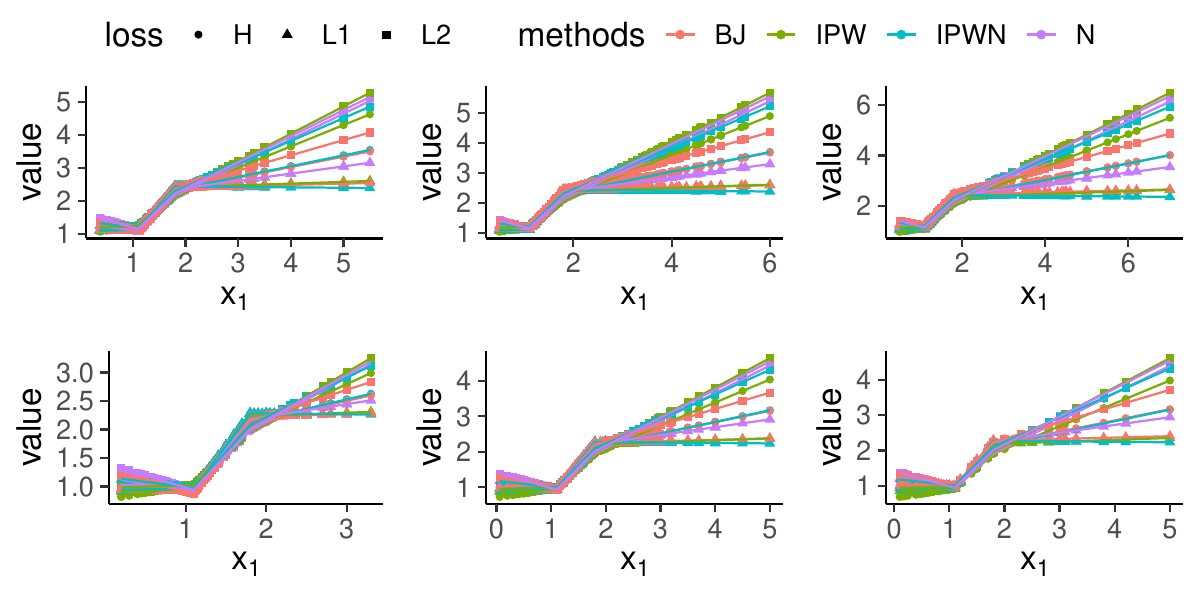}
    \vspace{-19mm}
    \caption{\it The predicted values $\hat{f}^*_n(X_i)$ versus its first argument $X_{i1}$ for $i=1,\ldots,n$, obtained from the proposed methods and the N method, with the second argument of $\hat{f}^*_n(X_i)$ fixed as 1, 2, or 3 (left to right columns) and with the third argument of $\hat{f}^*_n(X_i)$ fixed as 1 or 2 (top to bottom rows).
}
    \label{fig: 5}
\end{figure}

\begin{figure}[h]
    \centering
   \includegraphics[width=0.9\textwidth]{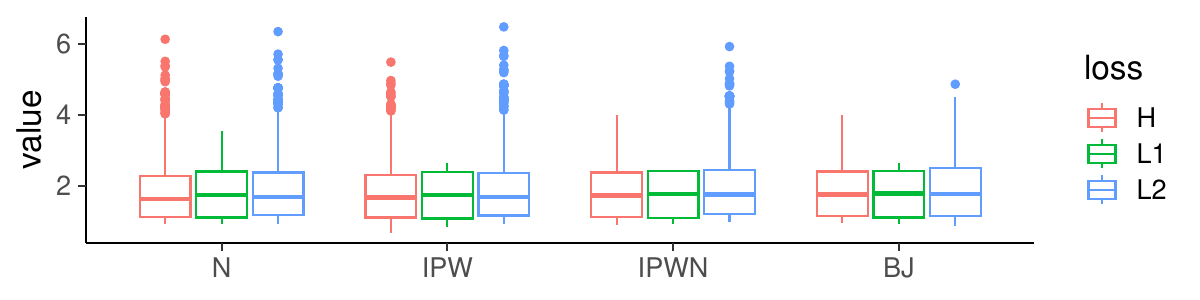}
   \vspace{-8mm}
    \caption{\it Boxplots of $\left\{\hat{f}^*_n(X_i):i=1,\ldots,n\right\}$.}
    \label{fig: 6}
\end{figure}

We analyze the data using the methods considered in Section \ref{subsec: am} except the R  method. Let $\hat{f}^*_n(X_i)$ denote the predicted value for the observed covariates $X_i$ with $i=1,\ldots,n$, obtained from each method. To visualize those estimates, for each estimate $\hat{f}^*_n(\cdot)$, we fix the second argument to be 1, 2, or 3, and the third argument to be 1 or 2, and then in Figure \ref{fig: 5}, we plot $\hat{f}^*_n(\cdot)$ against the first argument, denoted $X_1$. The results yielded from the proposed methods are similar and tend to deviate from those produced by the N  method at the left and right ends of the first argument's range. Additionally, in Figure \ref{fig: 6}, we plot boxplots for the predicted values $\left\{\hat{f}^*_n(X_i):i=1,\ldots,n\right\}$ for all considered methods.
\vspace{-3mm}
\vspace{-6mm}

\section{Discussion}\label{sec: d}
In this paper, we introduce two strategies -- the Buckley-James (BJ)-type adjustment and the IPW adjustment -- for modifying loss functions to mitigate the impact of MNAR data and develop a boosting prediction procedure. The BJ-type adjustment requires determining the conditional expectation of the loss function, which relies on knowledge of the conditional distribution of $Y$ given $X$ and $R=0$. In contrast, the IPW adjustment weights the loss function by $\pi(y,x)$, derived from modeling the conditional distribution of $R$ given $Y$ and $X$. Neither method is universally superior; their effectiveness depends on the feasibility of their respective modeling assumptions. When model reliability is uncertain, applying both methods to compare results can help assess sensitivity to assumptions. 

While our numerical studies focus on three loss functions: $L_1$, $L_2$, and the Huber loss, our theoretical results are applicable to other loss functions as well. The proposed strategies can be extended to accommodate other machine learning techniques, such as support vector machines, tree-based methods, and neural networks, which can be carried out by modifying their respective loss functions with the proposed schemes designed to handle MNAR data.

The validity of the proposed methods depends on certain conditions, as outlined in Section S.1 of the Supplementary Materials. These include identification conditions, the consistent estimation of key components such as $\pi(y,x)$ and/or $f(y|X=x,R=1)$, and standard conditions related to the loss function and covariates. These assumptions are important for establishing the theoretical guarantees of our methods. While some assumptions can be verified directly, many are challenging to validate in practice and may not always hold. Consequently, the practical performance of the proposed methods depends on how well these assumptions are satisfied in specific applications. 

Our primary objective is to identify an optimal function of covariates for predicting the response variable, rather than focusing on inference about the underlying model parameters. We prioritize predictive performance within the MNAR framework and do not directly address hypothesis testing or parameter estimation uncertainty.

\section*{Supplementary Materials}
\begin{description}
\item[Text document:] A .pdf file contains regularity conditions, proofs of Propositions \ref{prop1} and \ref{prop2}, Theorems \ref{thm1} and \ref{thm2},
and additional simulation results.
\item[R-code:] A .zip file, named ``code.zip'', includes R scripts for the simulations and data analyses presented in Sections \ref{sec: ss} and \ref{sec: da}.
\end{description}

\section*{Acknowledgments}
\noindent
The authors thank the review team for their comments on the initial version of the manuscript and Dr. Morikawa for providing the KLIPS data and the R code used in \cite{MorikawaKim2021}. Yi is a Canada Research Chair in Data Science (Tier 1). Her research is supported by the Natural Sciences and Engineering Research Council of Canada (NSERC) and the Canada Research Chairs Program.

\bibliographystyle{apalike}
\bibliography{main}
\end{document}